\documentclass[twocolumn,showpacs,preprintnumbers,amsmath,amssymb,aps,prb,superscriptaddress]{revtex4}
\usepackage{graphicx}

\usepackage{bm}
\usepackage[usenames]{color}
\usepackage{ulem}

\newcommand{\redsout}[1]{}
\newcommand{\red}[1]{#1}

\begin{document}

\title{Signature of the microcavity exciton polariton relaxation mechanism in
the polarization of emitted light}

\author{Georgios Roumpos}
	\email{roumpos@stanford.edu}
	\affiliation{E. L. Ginzton Laboratory, Stanford University, Stanford, CA, 94305, USA}
\author{Chih-Wei Lai}
	\affiliation{E. L. Ginzton Laboratory, Stanford University, Stanford, CA, 94305, USA}
	\affiliation{National Institute of Informatics, Hitotsubashi, Chiyoda-ku, Tokyo 101-8430, Japan}
\author{T. C. H. Liew}
    \affiliation{Centre for Quantum Technologies, National University of Singapore, Singapore 117543}
\author{Yuri G. Rubo}
    \affiliation{School of Physics and Astronomy, University of Southampton, Highfield, Southampton SO17 1BJ, UK}
    \affiliation{Centro de Investigaci\'on en Energ\'{\i}a, Universidad Nacional Aut\'onoma de M\'exico, Temixco, Morelos, 62580, Mexico}
\author{A. V. Kavokin}
    \affiliation{School of Physics and Astronomy, University of Southampton, Highfield, Southampton SO17 1BJ, UK}
    \affiliation{Marie-Curie Chair of Excellence ``Polariton devices'', University of Rome II, 1, via della Ricerca Scientifica, Rome, 00133, Italy}
\author{Yoshihisa Yamamoto}
	\affiliation{E. L. Ginzton Laboratory, Stanford University, Stanford, CA, 94305, USA}
	\affiliation{National Institute of Informatics, Hitotsubashi, Chiyoda-ku, Tokyo 101-8430, Japan}

\date{\today}


\begin{abstract}
We have performed real and momentum space spin-dependent spectroscopy of
spontaneously formed exciton polariton condensates for a non-resonant pumping
scheme.
Under linearly polarized pump, our results can be understood in terms of
spin-dependent Boltzmann equations in a two-state model.
This suggests that relaxation into the ground state occurs after
multiple phonon scattering events and only one polariton-polariton scattering.
For the circular pumping case, in which only excitons of one spin are injected,
a bottleneck effect is observed, implying inefficient relaxation.
\end{abstract}

\maketitle

\section{Introduction\label{sec:Intro}}

Bose-Einstein condensation (BEC) is an active field of research, especially
after its realization in dilute alkali
gases \cite{Anderson1995,Davis1995}.
Microcavity exciton polaritons \cite{Weisbuch1992,Yamamoto2000,Kavokin2003},
composite quasi-particles consisting of quantum well (QW) exciton and
microcavity photon components, have been
proposed as candidates for BEC \cite{Imamoglu1996}.
Due to their low mass, the critical temperature for BEC is expected to be
high, even up to room temperature \cite{Christopoulos2007}.
The confinement in two dimensions, along with the dual exciton-photon character
of polaritons, enables interesting optical studies.
Indeed, several characteristic signatures of dynamical condensation have been
reported in recent years \cite{HDeng2003,JKasprzak2006,HDeng2007}.

However, the lifetime of polaritons is short, on the order of $10$ psec
in our GaAs-based sample when condensation is observed, so the system is
inherently dynamical.
In previous studies, the final energy distribution of polaritons was compared to
the Bose-Einstein distribution for steady-state \cite{JKasprzak2006} or
time-resolved \cite{HDeng2006} data.
These results are explained by modeling the relaxation
mechanism in terms of polariton-acoustic phonon and polariton-polariton
scattering \cite{Tassone1999,Porras2002,Doan2005}.
However, taking into account the polariton spin degree of freedom introduces
further complications, due to the interplay between energy and spin
relaxation \cite{Kavokin2004,Shelykh2005,Cao2008}.

Here, we report the insights we gained on the relaxation mechanism, based on
polarization-dependent studies of exciton polariton condensation under
non-resonant incoherent pumping.
For linearly polarized pump, the condensate emission develops both non-zero
linear and circular polarization.
We observed rotation of the linear polarization axis by $\sim 90^\circ$ between
the pump and condensate.
The exact rotation angle is correlated with the handedness of the observed
circular polarization.
These signatures are similar to the observations of a
parametric oscillator experiment \cite{Krizhanovskii2006}, which were
interpreted \cite{Schumacher2007} in terms of spin-asymmetric
polariton-polariton interaction
\cite{Kuwata-Gonokami1997,Ciuti1998,Eastham2003}.
We use a two-state model employing the spin-dependent Boltzmann
equations \cite{Shelykh2005} to understand our experimental results.
The agreement we obtain reveals the similarities of the non-resonant pumping
scheme to parametric oscillator (magic angle) geometries \cite{Stevenson2000}.
In the former case, it is believed that polaritons suffer multiple scatterings
with phonons and other polaritons
before reaching the $k_x\sim 0$ region, so any phase coherence inherited from
the laser should be lost, whereas in the latter case only one
polariton-polariton interaction occurs \cite{Keeling2007}.
Further, the observed spectra under circular pumping, show a bottleneck effect.
This suggests that polaritons cannot efficiently relax into the ground state
when only one spin species is present.
A similar suppression of the scattering rate was observed in parametric
amplification experiments \cite{Tartakovskii2000,KavokinPRB2003}.

In Section \ref{sec::setup} we describe our experimental setup. Our measurements of the Stokes vector and the corresponding theoretical model are presented in Sections \ref{sec:Stokes} and \ref{sec::Theory} respectively. Section \ref{sec:Bottleneck} covers the relaxation bottleneck under circularly-polarized pumping. Our conclusions are drawn in Section \ref{sec:Conclusions}.
In the Appendix, we write down the equations used in our theoretical model.

\section{Experimental setup\label{sec::setup}}

\begin{figure}[t!]
\includegraphics[width=3.4in]{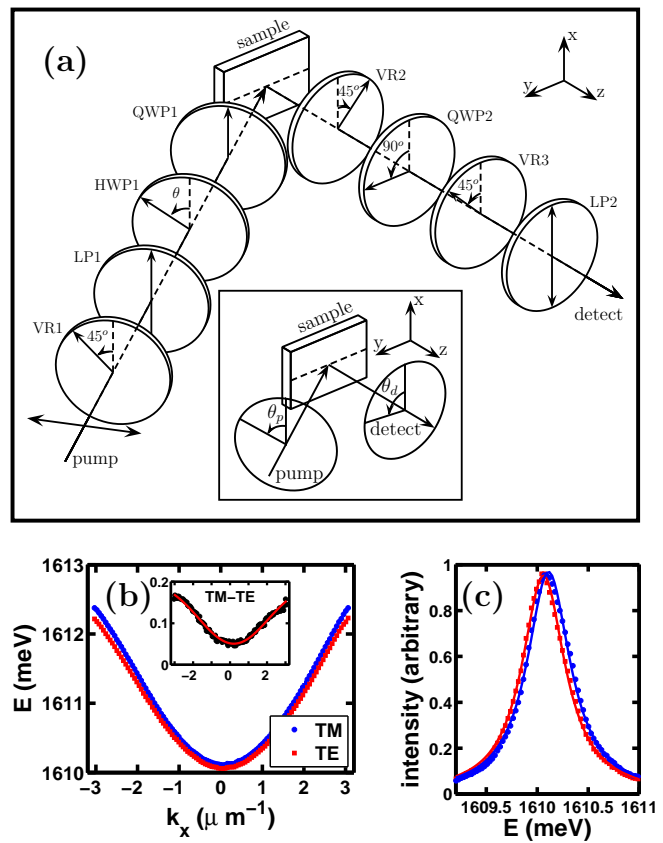}
\caption{(color online) (a) The polarization measurement setup.
Vectors label the fast or polarization axes of the optical components.
The laser pump is initially horizontally polarized ($\theta_p=90^o$), and is
incident at an angle of $55^o$ with respect to the growth direction $z$.
Luminescence is collected along the z-axis.
The first variable retarder (VR1) and linear polarizer (LP1) work as a variable attenuator.
By rotating a half waveplate (HWP1), and by using a removable quarter waveplate
(QWP1), we can implement various polarization states for the pump.
The second variable retarder (VR2) is used as a zero, half,
or quarter waveplate.
The combination of a quarter waveplate (QWP2), variable retarder (VR3) and
linear polarizer (LP2) is used for detection of a particular linear
polarization state, depending on the retardance of VR3.
Inset: Definition of angles $\theta_p$ and $\theta_d$
corresponding to the polarization axes of the pump and detection respectively.
(b) Measured dispersion curves for TM ($\theta_d=0^\circ$ - blue dots) and TE
($\theta_d=90^\circ$ - red squares) luminescence for low excitation density
($60\mu m^{-2}$ per pulse per QW).
The plotted points are the first moments of measured spectra for every $k_x$.
A small ground state splitting is visible.
$k_x=3\mu m^{-1}$ corresponds to $21^\circ$ in air.
Inset: The measured TM-TE splitting (black dots) and the theoretical prediction
for our sample parameters with a superimposed ground state splitting of
$50\mu eV$ (red line).
(c) Measured spectra for $k_x=0$ (points) fitted with Lorentzians (lines).
\label{fig:MeasurementGeometry}}
\end{figure}

The sample is the same as in Ref. \onlinecite{Utsunomiya2008}.
\red{
It consists of an ${\rm AlAs}$ $\frac{\lambda}{2}$ cavity sandwiched between two
distributed Bragg reflector (DBR) mirrors.
The upper and lower mirrors are made of 16 and 20 pairs respectively of
${\rm AlAs}$ and ${\rm Ga_{0.8}Al_{0.2}As}$.
3 stacks of 4 ${\rm GaAs}$ QW's are grown at the central three antinodes of the cavity.
}
The spectroscopy setup is described in Ref. \onlinecite{CWLai2007},
and it allows us to perform near field (NF - real space) and far field
(FF - momentum space) imaging and spectroscopy.
\red{
That is, we can measure energy-resolved luminescence as a function
of position or of in-plane momentum.
}
The measurements reported here are taken from a spot on the
sample with photon-exciton detuning $\delta=+6meV$, while the Rabi splitting
is $2\hbar\Omega_{Rabi}=14meV$.
The sample is kept at a temperature of $7-8K$ on the cold finger of a He flow
cryostat.
The system is pumped with a mode-locked Ti-Sapphire laser of $2$psec pulse width
and $76$MHz repetition rate focused on an ellipse of diameters
$50\mu m$ and $30\mu m$.
For FF data, luminescence is collected through an aperture at the first image
plane corresponding to a circular area of $30\mu m$ diameter on the sample.
The pumping laser is incident at an angle of $55^\circ$
(Fig. \ref{fig:MeasurementGeometry} inset, corresponding wavenumber
$k_y=-7\mu m^{-1}$), at the exciton resonance wavelength.
The setup employs liquid crystal polarization components
as shown in Fig. \ref{fig:MeasurementGeometry}(a).
We can pump with linear polarization of varying angle $\theta_p$, as well as
general elliptical polarization. The detection can be performed for linear
polarization of arbitrary angle $\theta_d$, or for right- and left-circular
polarization.

Using the transfer matrix method \cite{Yeh2005,Zhu1990} for exciton
inhomogeneous broadening of $3meV$, as measured at the far blue detuned
regime,
we estimate that the absorbed laser power for TM ($\theta_p=90^\circ$)
and TE ($\theta_p=0^\circ$) pumping is $\sim 4\%$ and $\sim 0.9\%$
respectively of the incident power.
We assume that the absorption efficiency is independent of power.
In the rest of the paper, the various pump polarization states refer to the
actually absorbed light inside the cavity, taking into account the calculated
differential absorption of TM and TE pumping.

A ground state ($k_{x,y}=0$) linear polarization splitting of $\sim 50\mu eV$,
similar to earlier studies
\cite{Krizhanovskii2006,Klopotowski2006,JKasprzak2007}, is measured for
low excitation power and the current sample orientation, (Fig.
\ref{fig:MeasurementGeometry}(b-c)) possibly due to crystal asymmetry
or strain.
The observed superimposed linear polarization splitting for $k_x\neq 0$ is
in quantitative agreement with a transfer matrix calculation (Fig.
\ref{fig:MeasurementGeometry}(b) inset).

\section{Stokes vector measurement\label{sec:Stokes}}

The polarization state of light is characterized by the following
three parameters (normalized with respect to the total power), which are
equivalent to the Stokes parameters as originally defined \cite{Hecht1970}:
\begin{equation}
S_1= \frac{I_{0^\circ}-I_{90^\circ}}{I_{0^\circ}+I_{90^\circ}},\quad
S_2= \frac{I_{45^\circ}-I_{-45^\circ}}{I_{45^\circ}+I_{-45^\circ}},\quad
S_3= \frac{I_{L}-I_{R}}{I_{L}+I_{R}}.
\label{eq:StokesParametersDefinition}
\end{equation}
$I_{0^\circ}$, $I_{90^\circ}$, $I_{45^\circ}$, and $I_{-45^\circ}$ are the
intensities of the linearly polarized components at $\theta_d=0^\circ$,
$90^\circ$, $45^\circ$, and $-45^\circ$ respectively. $I_L$ and $I_R$ are
the intensities of the left- and right-circularly polarized components
respectively.
From the above parameters, we can calculate the degree of linear polarization
(DOLP) and the angle of the major linear polarization axis $\psi$
\begin{equation}
\text{DOLP }=\sqrt{S_1^2 + S_2^2},\quad
\psi = \frac{1}{2} \arctan\left(\frac{S_2}{S_1}\right).
\label{eq:PolarizationParameters}
\end{equation}

\begin{figure}[t!]
\includegraphics[width=3.4in]{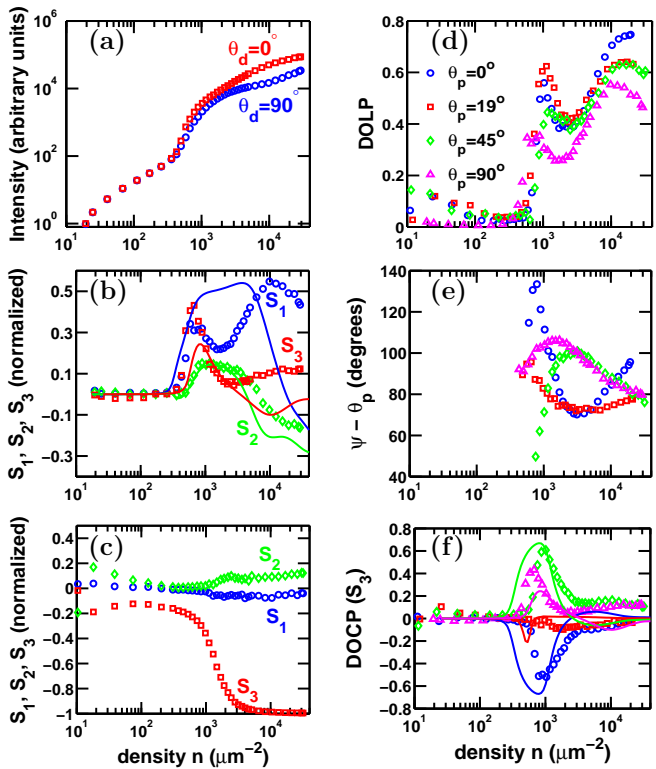}
\caption{(color online)
Measurement of Stokes parameters (markers) compared with the theoretical
model (solid lines).
(a) Horizontal pumping ($\theta_p=90^\circ$). Collected luminescence for
$\theta_d=0^\circ$ (red squares) and $\theta_d=90^\circ$ (blue circles)
vs. injected particle density in $\mu m^{-2}$ per pulse per QW.
A clear threshold is observed at $5\times 10^2\mu m^{-2}$.
(b-c) Degree of polarization measurement for (b) $\theta_p=90^\circ$
linear pumping and (c) left circularly polarized pumping.
Blue circles: $S_1$, green diamonds: $S_2$, red squares: $S_3$
defined in eq. (\ref{eq:StokesParametersDefinition}).
(d-f) Calculated polarization parameters from the measurement of the
Stokes parameters for linear pumping (eqs.
(\ref{eq:StokesParametersDefinition}-\ref{eq:PolarizationParameters})).
(d) DOLP.
(e) Angle for major axis of linear polarization $\psi$ relative to $\theta_p$
(f) Degree of circular polarization ($S_3$).
\label{fig:DOP}}
\end{figure}

We record the far field spectra for varying pumping power and polarization
angles $\theta_p$ and $\theta_d$, and sum the intensities inside the area
$\left| k_x\right|<0.55\mu m^{-1}$ (corresponding to $4^\circ$).
The observed normalized intensities $I_{\theta_d}$ are only weakly dependent on
the choice of this area, and are shown in Fig. \ref{fig:DOP}.
In Fig. \ref{fig:DOP}(a) we plot the measured luminescence intensity for
linearly polarized light along $\theta_d=0^\circ$ and $\theta_d=90^\circ$
as a function of pumping power in units of the generated polariton density
per pulse per QW. The pump is horizontally polarized ($\theta_p=90^\circ$).
The data show a non-linear increase above a threshold
density of $\sim 400\mu m^{-2}$, which marks the onset of condensation.
By measuring all six intensities required by equation
(\ref{eq:StokesParametersDefinition}), we calculate the three Stokes parameters.
The results for this pump polarization ($\theta_p=90^\circ$) are plotted in
Fig. \ref{fig:DOP}(b) along with the theoretical curves, to be discussed
in the next section.

For circularly polarized pump (Fig. \ref{fig:DOP}(c)) and well above threshold,
the signal is perfectly circularly polarized, up to $-99.4\%$.
This is due to the short polariton lifetime ($\sim 10ps$),
which is shorter than the spin relaxation time.
The negative sign of $S_3$ means that the angular momentum of the emitted
photons along the z-axis is the same as that of the optically injected
exciton polaritons, since we pump and detect from the same side of the sample
(Fig. \ref{fig:MeasurementGeometry}(a)).

\begin{figure}%[t!]
\includegraphics[width=3.4in]{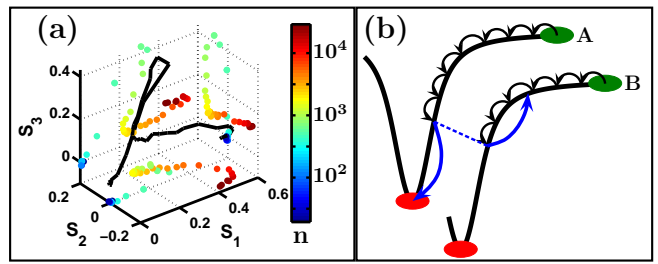}
\caption{(color online) (a) The path followed by the polarization vector for increasing
excitation power ($\theta_p =90^\circ$ linearly polarized pumping).
The projections on the three normal planes are shown with colored dots,
the color scale corresponding to the injected polariton density
$n$ in $\mu m^{-2}$ per pulse per quantum well (QW), as shown in the colorbar.
(b) Schematic of the proposed relaxation mechanism: the optically excited
polariton A, first looses energy by phonon scattering, then scatters with
one other polariton (B) and populates the ground state.
\label{fig:Poicare}}
\end{figure}

We next focus on linearly polarized pumping and vary the direction of linear
polarization for the pump ($\theta_p$).
Above threshold, a non-zero degree of
linear polarization develops (Fig. \ref{fig:DOP}(d)),
while the polarization direction is rotated by $\sim 90^\circ$ compared
to the pump (Fig. \ref{fig:DOP}(e)). Also, a circularly polarized component
emerges, with $S_3$ changing sign for varying $\theta_p$
(Fig. \ref{fig:DOP}(f)).
The sign change is correlated with the deviation of $\psi - \theta_p$ from
$90^\circ$.
The path followed by the polarization vector for increasing power and
$\theta_p=90^\circ$ linearly polarized pumping is plotted in Fig.
\ref{fig:Poicare}(a).

\section{Theoretical model\label{sec::Theory}}

To interpret these results we have used a simplified model based on
the spin-dependent Boltzmann equations for polaritons in
microcavities~\cite{Shelykh2005}. Our model is based upon two
states, representing the condensate and reservoir, each
characterized by a $2\times2$ spin density matrix.
The polariton-polariton scattering matrix element in parallel spin
configuration, $\alpha_1$ (positive), is believed to be much greater in
magnitude~\cite{Ciuti1998,Schumacher2007} than that in antiparallel
configuration, $\alpha_2$ (negative). Therefore, calculating the
transition rates we keep only terms $\propto\alpha_1^2$ and the
interference terms $\propto\alpha_1\alpha_2$.
We assume the reservoir is quickly populated
by the pump from fast polariton-phonon relaxation. Then we consider
the polariton-polariton scattering processes, which populate the
condensate (Fig. \ref{fig:Poicare}(b)).

The spin-anisotropy of the polariton-polariton
interactions gives rise to two important effects. First, a
$90^\circ$ rotation of the linear polarization appears upon
one polariton-polariton scattering, which has been evidenced
in parametric oscillator experiments in magic angle~\cite{Krizhanovskii2006}
as well as degenerate configurations~\cite{Leyder2007}.
This is because of the difference between the scattering matrix elements
of linearly polarized polaritons
\begin{align}
\left\langle\phi,\phi\right| V\left|\phi,\phi\right\rangle
&= \frac{1}{2} \left( \alpha_1 + \alpha_2 \right), \\
\left\langle\phi+90^\circ,\phi+90^\circ\right| V\left|\phi,\phi\right\rangle
&= \frac{1}{2} \left( \alpha_1 - \alpha_2 \right).
\end{align}
$V$ is the polariton-polariton interaction operator, and
$\left|\phi\right\rangle$ is the linear superposition
$\frac{1}{\sqrt{2}}\left( \left|\uparrow\right\rangle +e^{2i\phi}
\left|\downarrow\right\rangle \right)$ of spin-up and spin-down polaritons.
We note that if multiple polariton-polariton scattering events are
involved, the initial polarization information should be lost.

Second, if there is an
imbalance of the populations in the two spin components (in
either the condensate or the reservoir) then a self-induced Larmor
precession of the condensate and reservoir Stokes vector occurs.
This is because of the difference in the polariton-polariton interaction energy
between the different spin components.
This precession becomes faster by increasing the polariton population.
Therefore, at high pumping rates, the degree of linear polarization of the
luminescence decays in our time-integrated data (Fig. \ref{fig:DOP}(d)).

Other polarization sensitivity derives
from an assumed energy splitting between states linearly
polarized at $19^\circ$ and $109^\circ$,
as is evidenced from Fig. \ref{fig:MeasurementGeometry}(c) and from the lack
of circularly polarized component in the luminescence for excitation with
$\theta_p=19^\circ$.
This splitting causes a rotation of the Stokes vector if the
reservoir state is not an eigenstate with linear polarization of
$19^\circ$ or $109^\circ$, which results in non-zero $S_3$
(Fig. \ref{fig:DOP}(f)).
The condensate Stokes
parameters are time integrated and normalized by the time integrated
condensate population for comparison to the experimental results.

The results of our model are represented by solid lines in
Fig.~\ref{fig:DOP}. We assumed a
condensate lifetime of $2ps$, reservoir lifetime of $100ps$, pulse
duration of $2ps$, $\alpha_2/\alpha_1$=$-0.025$, and polarization
splittings of $50\mu eV$ for both the condensate and reservoir.
The final equations and the value we used for $\alpha_1$ are provided
in the Appendix.  The main features of our experimental results are
explained within this model.

\section{Relaxation bottleneck under circularly-polarized
pumping\label{sec:Bottleneck}}

\begin{figure}[t!]
\includegraphics[width=3.4in]{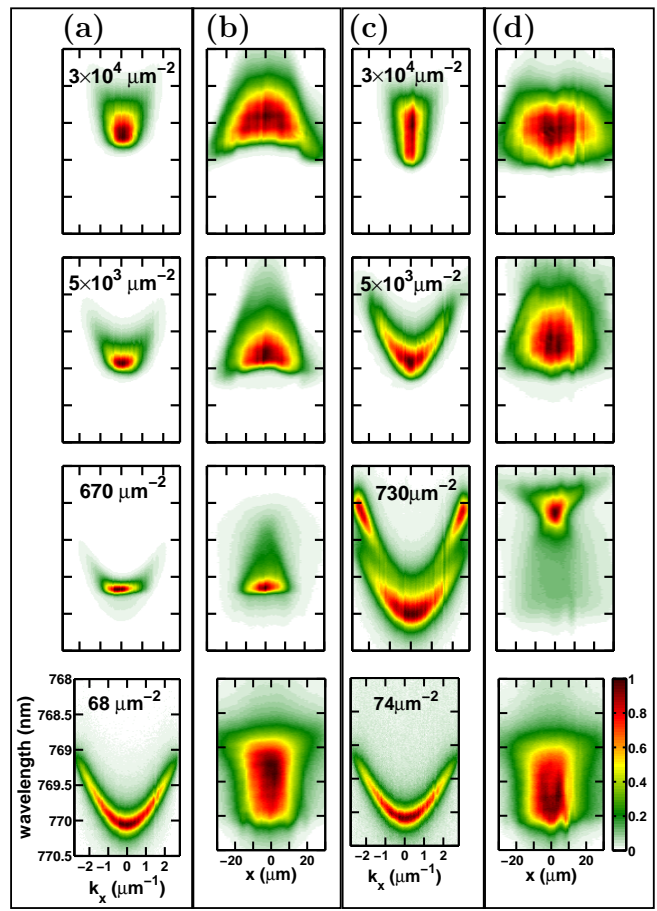}
\caption{(color online) Far-field (FF) ($k_x$ in $\mu m^{-1}$ vs. wavelength in $nm$)
and near-field (NF) ($x$ in $\mu m$ vs. wavelength in $nm$) spectra for various
injected particle densities (in $\mu m^{-2}$ per pulse per QW).
(a) FF, $\theta_p=90^\circ$, $\theta_d=0^\circ$.
(b) NF, same pumping-detection scheme.
(c) FF, left circular pump, right circular detection.
Note that the projection of angular momentum along the z-axis (Fig.
\ref{fig:MeasurementGeometry}(a)) has the same sign for both pump and detected
photons.
(d) NF, same pumping-detection scheme.
\label{fig:Spectra}}
\end{figure}

In Fig. \ref{fig:Spectra} we compare the FF and NF spectra for
two pumping schemes, namely linear
($\theta_p=90^\circ$, Fig. \ref{fig:Spectra}(a-b)) and left-circular
(Fig. \ref{fig:Spectra}(c-d)) polarizations.
Under linear pumping, we observe that the linewidth narrows at threshold, and
luminescence is concentrated around $k_x=0$ and $x=0$.
For higher excitation power, the momentum and position distributions broaden
and the condensate energy blue-shifts.
Under circular pumping and at just above threshold,
relaxation bottleneck is observed in momentum space at $k_x\sim\pm 2.3\mu
m^{-1}$ ($\pm 16^\circ$ in air), while in real space luminescence is
concentrated at the center of the excitation spot.
This implies that relaxation into the zero momentum region is only efficient
when both spin species are present.
For higher excitation power, luminescence is mainly observed around $k_x=0$
and $x=0$, similar to the linear pumping case.
This result is consistent with previous parametric amplification experiments
\cite{Tartakovskii2000,KavokinPRB2003}, where a suppression of the scattering
rate towards the zero-momentum region was observed when only one spin species
was present.
%\red{
%Our simple two-state model presented in the last section does not allow
%a consistent interpretation of this data along with the data presented in
%Section \ref{sec:Stokes}.
%}

\red{
Polariton condensation is a competition between relaxation and decay from the cavity.
Our data suggest that relaxation is more efficient in the linearly polarized pump case,
whereas decay is more efficient in the circularly polarized pump case.
On the other hand, our simple two-state model treats the relaxation rate as a free parameter.
Derivation of this rate involves a full many-body calculation,
where all states in momentum space need to be considered.
A more sophisticated model is therefore needed to understand the results of this section.
}

\begin{figure}[t!]
\includegraphics[width=3.4in]{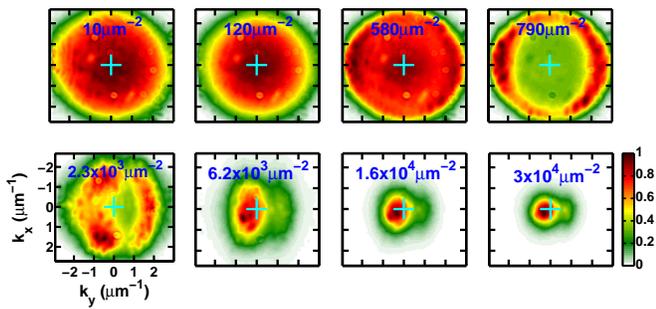}
\caption{(color online) Momentum space images for left-circularly polarized pumping and
right-circularly polarized detection (same scheme as in Fig.
\ref{fig:Spectra}(c-d)). For increasing pumping power, a ring pattern
develops and the images lose reflection symmetry.
The cyan crosses mark the origin in each figure.
The pump is incident at $\left( k_x,k_y\right)=(0,-7)\mu m^{-1}$.
\label{fig:FFImages}}
\end{figure}

\begin{figure}[t!]
\includegraphics[width=3.4in]{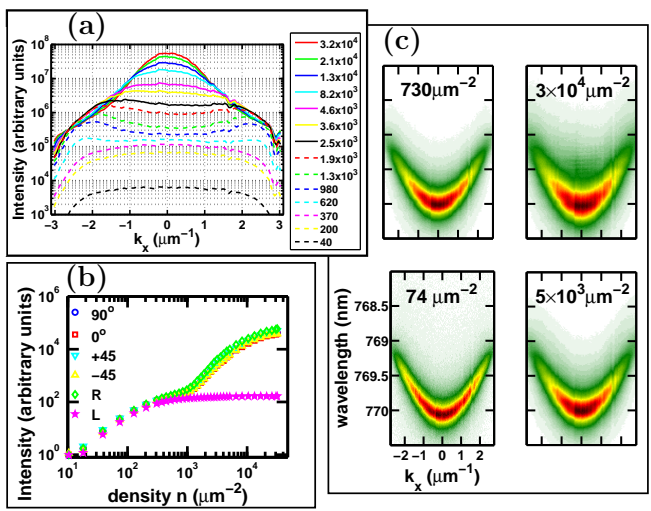}
\caption{(color online) (a) The momentum space distribution along the x-axis for
various polariton densities $n$ (in $\mu m^{-2}$ per pulse per QW).
The pump is left-circularly polarized, and the detection right-circularly
polarized (same scheme as in Figs. \ref{fig:Spectra}(c-d) and
\ref{fig:FFImages}).
At $n\sim 600\mu m^{-2}$ two peaks appear around $k_x=\pm 2.3\mu m^{-1}$,
which move towards $k_x=0\mu m^{-1}$ for increasing $n$.
Eventually, a central peak appears and dominates the luminescence.
(b) Luminescence inside the area $\left| k_x\right|<0.55\mu m^{-1}$ for the six
different polarization states of eq. \ref{eq:StokesParametersDefinition} as a
function of polariton density under left-circularly polarized pumping.
A stimulation threshold is observed at $n\sim 10^3\mu m^{-2}$.
(c) Far-field (FF) spectra for left-circularly polarized detection
(represented by magenta stars in (b)) for various pumping powers.
A broad distribution following the lower polariton dispersion is always
observed.
\label{fig:CoPolarizedData}}
\end{figure}

The inefficient cooling for the circular pumping case is further evidenced
in the FF images presented in Fig. \ref{fig:FFImages} for various pumping
powers. Above threshold, they do not possess the
$k_y\leftrightarrow -k_y$ reflection symmetry.
\red{
The laser pump is incident at $\left( k_x,k_y\right)=(0,-7)\mu m^{-1}$,
so the polariton distribution is shifted towards the source.
On the contrary, under linearly polarized pumping the momentum space
distribution is always spherically symmetric.
}
Detailed data of the momentum space distribution along the $x-$axis for
increasing pumping power are shown in Fig. \ref{fig:CoPolarizedData}(a).
The cross-circularly polarized component is much weaker above a threshold
pumping power, as shown in Fig. \ref{fig:CoPolarizedData}(b)),
and does not condense (Fig. \ref{fig:CoPolarizedData}(c)).

\begin{figure}[t!]
\includegraphics[width=3.4in]{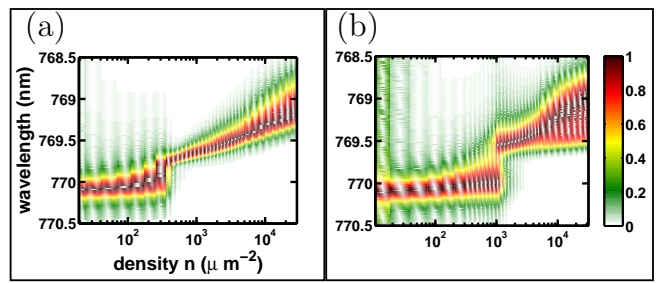}
\caption{(color online) (a) The measured spectra near zero momentum
($\left| k_x\right|<0.55\mu m^{-1}$) for linearly polarized pumping
($\theta_p=90^\circ$, $\theta_d=0^\circ$) as a function of polariton density.
(b) Same spectra for left-circularly polarized pump and right-circularly
polarized detection.
\label{fig:EnergyShift}}
\end{figure}

Fig. \ref{fig:EnergyShift}(a) shows the measured spectra near zero momentum
($\left| k_x\right|<0.55\mu m^{-1}$) for linearly polarized pumping
($\theta_p=90^\circ$, $\theta_d=0^\circ$).
We observe a linewidth decrease and blue shift just above threshold.
We note that the observed energy shift shows an almost logarithmic increase
as a function of pumping power, similar to Ref. \onlinecite{Bajoni2008}.
From a polariton-polariton interaction point of view, a linear increase would
be expected.
Fig. \ref{fig:EnergyShift}(b) shows the same spectra for left-circularly
polarized pump and right-circularly polarized detection.
We observe a similar blue shift, but no linewidth narrowing.
\red{
The reason for the different spectral linewidths is not well understood.
It might indicate that the temporal coherence is not necessarily enhanced
with increasing accumulation of polaritons near the zero in-plane momentum.
}

\section{Conclusions\label{sec:Conclusions}}

In conclusion, we studied polarization-dependent luminescence from an exciton
polariton system as a function of pump power and polarization in a
non-resonant pumping geometry.
Spin-dependent polariton-polariton interaction manifests itself in the
rotation of the linear polarization axis by $\sim 90^\circ$ under linearly
polarized pumping.
This can be understood in terms of a two-state model,
suggesting that polaritons populate the condensate after multiple
phonon scatterings and only one polariton-polariton scattering.
In addition, when only one spin species is injected, we observed a relaxation
bottleneck. This phenomenon is typically attributed to inefficient
relaxation, leading to photon leakage from the cavity before polaritons reach
the zero-momentum region.
Full determination of the polarization of polariton condensates
reveals that the spin degree of freedom plays an important role in
understanding the relaxation mechanism of microcavity exciton polaritons.

\begin{acknowledgments}
G.R. acknowledges support from JST/SORST and Special Coordination
Funds for Promoting Science and Technology.
T.C.H.L., Y.G.R., and A.V.K. would like to thank E.P.S.R.C.
for financial support.
A.V.K. thanks Ivan Shelykh for useful comments.
\end{acknowledgments}

\appendix*
\section{}
Here we present the equations used in the theoretical model of Section
\ref{sec::Theory}. The approach we have taken is based on the spin-dependent
Boltzmann equations for exciton-polaritons in microcavities of
Ref.~\onlinecite{Shelykh2005}.
We have considered two states, reservoir and condensate, each characterised by
a $2\times 2$ spin density matrix
\begin{eqnarray}
\left[
\begin{array}{cc}
R_\uparrow & \left(R_x-iR_y\right)\\
\left(R_x+iR_y\right) & R_\downarrow
\end{array}
\right],\nonumber\\
\left[
\begin{array}{cc}
N_\uparrow & \left(S_x-iS_y\right)\\
\left(S_x+iS_y\right) & N_\downarrow
\end{array}
\right].
\end{eqnarray}
Here $R_\uparrow$ and $R_\downarrow$ are the reservoir populations for spin-up and
spin-down polaritons, $R_x$ and $R_y$ are the pseudospin components that characterize 
the linear polarization degree measured in the horizontal-vertical and diagonal basis, respectively.
The circularly polarized component $R_z$ of reservoir pseudospin is 
$R_z=(R_\uparrow - R_\downarrow)/2$.
The corresponding numbers for the condensate are given by $N_\uparrow$,
$N_\downarrow$, $S_x$, $S_y$, and $S_z=(N_\uparrow - N_\downarrow)/2$.
$P_\uparrow$, $P_\downarrow$, $P_x$, and $P_y$ describe the pump.
For example, for TE pumping ($\theta_p=0^\circ$), we have
$P_\uparrow=P_\downarrow=P_x$.
The full rate equations we used are as follows,
\begin{eqnarray}
\frac{{\rm d}N_\uparrow}{{\rm d}t} =
-\Gamma N_\uparrow +(\omega_xS_y - \omega_yS_x)
+ W R_\uparrow^2\left( N_\uparrow +1\right),\\
\frac{{\rm d}N_\downarrow}{{\rm d}t} =
-\Gamma N_\downarrow -(\omega_xS_y - \omega_yS_x)
+W R_\downarrow^2\left( N_\downarrow +1\right),\\
\frac{{\rm d}S_x}{{\rm d}t} =
-\Gamma S_x +\omega_yS_z
-\frac{(\alpha_1-\alpha_2)}{\hbar}\left(S_z+R_z\right)S_y
\nonumber\\
+\frac{W}{2}\left(R_\uparrow^2+R_\downarrow^2\right)S_x
\nonumber\\
+\frac{W}{2}\frac{\alpha_2}{\alpha_1}\left(R_\uparrow+R_\downarrow\right)
\left(N_\uparrow+N_\downarrow+2\right)R_x,
\end{eqnarray}
\begin{eqnarray}
\frac{{\rm d}S_y}{{\rm d}t} =
-\Gamma S_y -\omega_xS_z
+\frac{(\alpha_1-\alpha_2)}{\hbar}\left(S_z+R_z\right)S_x
\nonumber\\
+\frac{W}{2}\left(R_\uparrow^2+R_\downarrow^2\right)S_y
\nonumber\\
+\frac{W}{2}\frac{\alpha_2}{\alpha_1}\left(R_\uparrow+R_\downarrow\right)
\left(N_\uparrow+N_\downarrow+2\right)R_y,\\
\frac{{\rm d}R_\uparrow}{{\rm d}t} = -\gamma R_\uparrow
+(\Omega_xR_y -\Omega_yR_x)
-W R_\uparrow^2\left( N_\uparrow+1\right)\nonumber\\
+P_\uparrow,\\
\frac{{\rm d}R_\downarrow}{{\rm d}t} = -\gamma R_\downarrow
-(\Omega_xR_y -\Omega_yR_x)
-W R_\downarrow^2\left( N_\downarrow+1\right)\nonumber\\
+P_\downarrow,\\
% \end{eqnarray}
%
% \begin{eqnarray}
\frac{{\rm d}R_x}{{\rm d}t} = -\gamma R_x +\Omega_yR_z
-\frac{(\alpha_1-\alpha_2)}{\hbar}\left(S_z+R_z\right)R_y\nonumber\\
-\frac{W}{2}\left[\left(N_\uparrow+1\right)R_\uparrow
+\left(N_\downarrow+1\right)R_\downarrow\right]R_x +P_x,\\
\frac{{\rm d}R_y}{{\rm d}t} = -\gamma R_y -\Omega_xR_z
+\frac{(\alpha_1-\alpha_2)}{\hbar}\left(S_z+R_z\right)R_x\nonumber\\
-\frac{W}{2}\left[\left(N_\uparrow+1\right)R_\uparrow
+\left(N_\downarrow+1\right)R_\downarrow\right]R_y +P_y.
\end{eqnarray}
Here 
$\omega_{x,y}$ and $\Omega_{x,y}$ are the Larmor frequencies corresponding
to the effective magnetic field due to the polarization splitting.
$\omega_{x,y}$ refer to the condensate and $\Omega_{x,y}$ refer to the
reservoir.
$\Gamma$ and $\gamma$ are the decay rates for the condensate and reservoir, respectively.
As discussed in Section \ref{sec::Theory}, we use the values
\begin{eqnarray}
\omega_x=\Omega_x=\frac{50 {\rm \mu eV}}{\hbar}\cos\left( 2\times 19^\circ\right),\nonumber\\
\omega_y=\Omega_y=\frac{50 {\rm \mu eV}}{\hbar}\sin\left( 2\times 19^\circ\right),\nonumber\\
\Gamma = 0.5\,{\rm ps}^{-1},\quad \gamma = 0.01\,{\rm ps}^{-1}.
\end{eqnarray}

The scattering rate from the reservoir to condensate is
$W=(2\pi/\hbar)\alpha_1^2\rho_i$, where
$\rho_i$ is the density of polariton states at the idler energy.
The idler energy is $E_i=2E_r-E_c$,  where $E_r$ and $E_c$ are the
energies of polaritons in the reservoir and condensate, respectively. 

We have used the value of $\alpha_1=5\times 10^{-4}\,\mathrm{meV}$,
which is the estimate of the interaction energy of 
two polaritons inside the excitation spot of 10$\,\mu m$ radius.
The scattering rate is estimated as $W=5\times 10^{-7}\,\mathrm{ps}^{-1}$.

\end{document}